\documentclass[3p]{elsarticle}
\usepackage{lineno}
\usepackage{mathptmx,amsmath,amssymb,bm}
\usepackage{float}
\usepackage[dvipsnames]{xcolor}
\usepackage{tikz}
\usepackage{graphicx,mwe}
 \usepackage{setspace}
\emergencystretch=\maxdimen
\def\firstpage{1}
\usetikzlibrary{arrows,shapes,automata,backgrounds,petri,positioning}
\usetikzlibrary{decorations.pathmorphing}
\usetikzlibrary{decorations.shapes}
\usetikzlibrary{decorations.text}
\usetikzlibrary{decorations.fractals}
\usetikzlibrary{decorations.footprints}
\usetikzlibrary{shadows}
\usetikzlibrary{calc}
\usetikzlibrary{spy}
\usetikzlibrary{calc,patterns,decorations.pathmorphing,decorations.markings}
\tikzstyle{spring}=[thick,decorate,decoration={zigzag,pre length=0.3cm,post length=0.3cm,segment length=3}]
\tikzstyle{small_spring}=[thin,decorate,decoration={zigzag,pre length=0.01cm,post length=0.01cm,segment length=1}]

\tikzstyle{damper}=[thick,decoration={markings,  
  mark connection node=dmp,
  mark=at position 0.5 with 
  {
    \node (dmp) [thick,inner sep=0pt,transform shape,rotate=-90,minimum width=15pt,minimum height=3pt,draw=none] {};
    \draw [thick] ($(dmp.north east)+(2pt,0)$) -- (dmp.south east) -- (dmp.south west) -- ($(dmp.north west)+(2pt,0)$);
    \draw [thick] ($(dmp.north)+(0,-5pt)$) -- ($(dmp.north)+(0,5pt)$);
  }
}, decorate]
\tikzstyle{process} = [rectangle, minimum width=3cm, minimum height=2.5cm, text 
centered, text width=3cm, draw=black, fill=orange!30]
\tikzstyle{region} = [rectangle, minimum width=3.0cm, minimum height=2.5cm, 
text centered, draw=black, fill=white!30]
\tikzstyle{arrow} = [thick,->,>=stealth]

\usepackage[numbers]{natbib}
\usepackage[caption=false]{subfig}
\title{A co-located partitions strategy for parallel CFD--DEM couplings} 

\author[rvt]{Gabriele Pozzetti\corref{cor1}, Xavier Besseron\corref{cor1}, Alban Rousset\corref{cor1}, Bernhard Peters\corref{cor1}}%\fnref{fn1}}
\ead{gabriele.pozzetti@uni.lu}
\cortext[cor1]{Principal corresponding author. \textit{Email:}  
gabriele.pozzetti@uni.lu, pozzetti.gabriele@gmail.com}
\address[rvt]{Campus Belval, Universit\'e du Luxembourg
6, Avenue de la Fonte, L-4364 Esch-sur-Alzette  Luxembourg}
\begin{document}

%\titlefigurecaption{{\large \bf \rm Applied Mathematics \& Information Sciences }\\ {\it\small An International Journal}}

% \title{On the performance of an overlapping-domain parallelization strategy for Eulerian-Lagrangian Multiphysics software}
% 
% \author{
% 	Gabriele Pozzetti\hyperlink{author1}{$^1$} 
% and Xavier Besseron\hyperlink{author2}{$^1$}
% and Alban Rousset\hyperlink{author3}{$^1$} 
% and Abdoul Wahid Mainassara Checkaraou\hyperlink{author4}{$^1$}
% and Bernhard Peters\hyperlink{author5}{$^1$}
% }
% \institute{$^1$LuXDEM Research Centre, Computational Engineering Department, University of Luxembourg}
% 
% 
% \titlerunning{On the performance of an overlapping-domain parallelization strategy for Eulerian-Lagrangian Multiphysics software}
% \authorrunning{G. Pozzetti et al.}
% 
% %corresponding author email
% \mail{gabriele.pozzetti@uni.lu}
% 
% \received{...}
% \revised{...}
% \accepted{...}
% \published{...}

\begin{abstract}
In this work, a new partition-collocation strategy for the parallel execution of CFD--DEM couplings is investigated.
Having a good parallel performance is a key issue for an Eulerian-Lagrangian software that aims to be
applied to solve industrially significant problems, as the computational cost of these couplings is 
one of their main drawback.
The approach presented here consists in co-locating the overlapping parts of the simulation domain of each software  on the same MPI process, 
in order to reduce the cost of the data exchanges.
It is shown how this strategy allows reducing memory consumption and inter-process communication between CFD and DEM
to a minimum and therefore to overcome an important parallelization bottleneck identified in the literature.
Three benchmarks are proposed to assess the consistency and scalability of this approach. 
A coupled execution on 280 cores shows that less than 0.1\% of the time is used to perform inter-physics data exchange.
\end{abstract}

\maketitle

\section{Introduction}\label{Introduction}
%\linenumbers
Eulerian-Lagrangian couplings are nowadays widely used to address engineering and 
technical problems.
In particular, CFD--DEM couplings have been successfully applied to study several
configurations ranging from 
mechanical~\cite{PozzettiPowdermet,BPAdditiveManufacturing}, to
chemical~\cite{Mahmoudi2016358} and environmental~\cite{NAG:NAG2387} engineering.

CFD--DEM coupled simulations are normally very computationally intensive, and, as already pointed out in~\cite{KafuiParallel},
the execution time represents
a major issue for the applicability of this numerical approach to complex scenarios.
Therefore, optimizing the parallel performance of such a coupling is a fundamental 
step for allowing large-scale numerical solutions of industrial and technical problems.

The parallelization of Eulerian-Lagrangian software is, however, rather delicate.
This is mainly due to the fact that the optimal partitioning strategies for those 
wireframes are different, and the memory requirement of a coupled solution can 
represent a major performance issue. 

Furthermore, since the coupling normally affects an extended domain region (often the whole computational domain),
the amount of information that is required to be exchanged is normally important. For this reason, 
highly efficient coupling approaches for boundary problems, like the one proposed in~\cite{Buis2006} may suffer 
for the extensive communication layer. At the same time, due to the Eulerian-Lagrangian nature of the coupling,
mesh-based communication as the one proposed in~\cite{joppich2006mpcci} cannot, by themselves take care of the 
information exchange.

One of the earliest attempt to parallelize a DEM algorithm was proposed in~\cite{WashingtonDEMPar},
where the authors distributed inter-particle contacts among processors on a machine featuring 512 cores.
In their scheme, all particle data was stored in every process resulting in a memory-intensive
computation that leads to a speedup of 8.73 for 512 cores with a 1672 particles assembly.
A later work~\cite{Maknickas2006} showed how by reducing the DEM inter-process communication,
a speedup of $\sim$11 within  16 process-computation of 100k particles was obtainable.
This proved how, for the sole Lagrangian software, the memory usage and the inter-process communication
can profoundly affect the parallel performance of the code.

The problem of memory consumption and inter-process communication becomes even more important 
when a Lagrangian code is coupled with an Eulerian one as the 
single pieces of software 
generally reach the optimum performance with different load and partitioning 
strategies.
This can lead to massive inter-process communication that can deeply affect the 
performance of the overall algorithm.
In~\cite{DarmanaParallel}, the authors proposed a mirror domain method for ensuring the 
correct passage of information between Lagrangian bubbles and an underlying fluid flow in a parallel execution.
This method consists in distributing global knowledge of all the Eulerian and Lagrangian domains, by 
letting a process performing only a part of the computation, and then distributing the information via gather-scatter operations.
It was shown in~\cite{DarmanaParallel}, how this strategy allows a correct access to information and an almost linear 
speedup up to 4 processes for a case with 648000 CFD cells and 100000 particles.
Nevertheless, the strategy wasn't able to offer significant advantages when operating on more than 32 processes. Therefore, 
 its application is limited to medium-small scale problems.
 
In order to cope with this problem, in~\cite{KafuiParallel}, the authors proposed a parallelization 
strategy based on a Jostle graph for the DEM part and an independent parallelization for the CFD part. 
This allowed reducing memory consumption while keeping an high inter-physics communication load.
The resulting coupling was observed to scale better than what previously seen in the literature
obtaining a speedup of $\sim$35 for 64 processes. Nevertheless, the coupled code was shown to perform 
significantly worse than the sole DEM part, proving how the inter-physics communication between the 
Eulerian and the Lagrangian part can induce significant performance issues.

This issue was underlined by~\cite{EfficientOpenMPICFD-DEM}, where a communication strategy based 
on non-distributed memory was implemented in order to reduce the communication costs. The results proved 
how this strategy can indeed reduce the DEM and the inter-physics communication, yet its advantages are limited
to the usage of $\sim$30 processes though not allowing very-large-scale simulations. 

An important attempt to study large-scale problems with an Eulerian-Lagrangian coupling was presented in 
~\cite{SediFoam}.
In this contributions, the authors propose a coupling between a highly efficient code for molecular dynamics 
and a well-known open-source software for CFD. The proposed decomposition strategy for CFD and molecular dynamics code are completely independent from 
each other, and all the required data for the inter-physics exchange is provided through a complex communication layer.
It is shown how the specific couple of codes can tackle large-scale problems and scale over more than one hundred of processes.
Nevertheless, when a large number of processes is used, the inter-physics communication becomes very important, taking up to 
30\% of the computation time.

Nowadays, there is still no reference solution for the parallel execution of CFD--DEM couplings, 
and even though several strategies have been proposed, they have been focusing 
on specific software. The aim of the current work is to investigate the performance of a 
parallelization strategy for a CFD--DEM four-way coupling that aims to minimize the memory consumption and the 
inter-physics communication.

The article is structured as follows.
Firstly, the Lagrangian and Eulerian parts of this coupling are presented, and 
the systems of equations solved by the DEM and CFD part described.
Secondly, the partitioning strategy based on co-located partitions is introduced.
Finally, three benchmark cases are proposed in order to assess the
consistency and performances of the proposed parallelization strategy.

In this work, we refer to the coupling between the XDEM 
platform~\cite{PozzettiIJMF,PozzettiICNAAM2016}, which is 
used to treat the Lagrangian entities, 
and the OpenFOAM~\cite{OpenFOAM} libraries, which is used to resolve the fluid flow
equations in a Eulerian reference. The parallelization strategy is presented in its general way 
and can be implemented for a generic CFD--DEM coupling independently from the specific software.

To the best of the authors' knowledge, the partitioning strategy for CFD--DEM couplings here proposed is the first one 
that succeeds in keeping the inter-physics communication negligible while running over hundreds of process.
\section{Methodology}

\begin{table*}[htbp]
\centering
\caption{List of variables}
\label{SimVar}
\begin{tabular}{p{1cm}p{5cm}p{1cm}p{5cm}}
%\arrayrulecolor{black}
\cline{1-4}
Symbol &  Variable & Symbol &  Variable  \\
\cline{1-4}
\\
 $\mathbf{u}_f$ &  fluid velocity &   $p$   & pressure \\ 
 $\mu_f$ &  fluid viscosity & $\mathbf{u}_c$ &  compression velocity\\
  $\mathbf{F_{b}}$ &  body force & $C$ &  surface curvature\\
  $\mathbf{F_{fpi}}$ &  fluid-particle interaction force & $\bm{\sigma}_f$ & fluid stress tensor \\
  $\rho_{f}$ &  fluid density & $\rho_p$ & particle density \\
  $A_p$ &  particle surface area & $\epsilon$ & local porosity \\
\cline{1-4} \\
 $\mathbf{M}_{coll}$ &  torque acting on a particle due to collision events &$\mathbf{F}_g$ &  gravitational force\\
 $m$ &  particle mass &$\mathbf{M}_{ext}$ &  external torque acting on a particle\\
 $\mathbf{I}$ &  particle moment of inertia & $\mathbf{u}$ &  particle velocity\\
 $\mathbf{F}_{coll}$ &  contact force acting on a particle   \\ 
 $\mathbf{F}_{ij}$ &  contact force acting on the particle $i$ due to the collision with particle $j$ & $\bm{\phi}$ &  particle orientation\\
 $\mathbf{F}_{drag}$ &  force rising from the particle-fluid interaction &$\bm{\omega}$ & particle angular velocity \\ 
\cline{1-4} 
\end{tabular}
\end{table*}

\subsection{Discrete Element Method (DEM)}

The Discrete Element Method (DEM) is a well-established approach for granular 
flows~\citep{Herrmann1995}. 
The XDEM platform~\cite{PozzettiIJMF,BaniasadiMelting,BPAdditiveManufacturing} aims 
to extend the application of the DEM by attaching chemical/thermodynamical 
variables on the particles,
and coupling the DEM core with codes for computational fluid dynamics (CFD).
More in general, the coupling between CFD and DEM is a common strategy for 
addressing industrially 
relevant problems~\citep{BPAdditiveManufacturing,Mahmoudi2016358}.
In this work, the DEM module of the XDEM platform is used to evolve a set of 
particles describing 
physical entities in the presence of a multiphase fluid.
Defining with $\mathbf{x}_i$ the positions, $m_i$ the masses, and $ \bm{\phi}$ the orientations, one can write
\begin{eqnarray}\label{DEM_EQUATIONS}
m_i \frac{{d}^2}{{dt}^2}\mathbf{x}_i & =& 
\mathbf{F}_{coll}+\mathbf{F}_{drag}+\mathbf
{F}_g , \\
\mathbf{I}_i \frac{{d}^2}{{dt}^2} \bm{\phi}_i & =& 
\mathbf{M}_{coll} + \mathbf{M}_{ext},
\end{eqnarray}
In equation \ref{DEM_EQUATIONS}, 
the term $\mathbf{F}_c$  indicates the collision force,
\begin{eqnarray}\label{FORCES}
\mathbf{F}_{coll} = \sum_{i\neq j} \mathbf{F}_{ij}( \mathbf{x}_j, \mathbf{u}_j, 
\bm{\phi}_j, \bm{\omega}_j ),
\end{eqnarray}
{with  $\mathbf{u}_j$ the velocity of particle $j$, and $\bm{\omega}$ the angular 
velocity.
The term  $ \mathbf{M}_{coll} $ indicates the torque acting on  
the particle due to collision{s}
\begin{eqnarray}\label{TORQUES}
\mathbf{M}_{coll} = \sum_{i\neq j} \mathbf{M}_{ij}(\mathbf{x}_j, \mathbf{u}_j, 
\bm{\phi}_j, \bm{\omega}_j ),
\end{eqnarray} 
with $\mathbf{M}_{ij} $ the torque acted from particle $j$ to particle $i$.
The term $\mathbf{F}_{drag}$ takes into account the force rising from the 
interaction with the fluid, 
and $\mathbf{F}_g$ corresponds to the gravitational force.
For what concerns $\mathbf{F}_{drag}$ the analytical expression of the term can 
be obtained from the literature. 
Many authors have focused on its description~\citep{DU20061401,AYENI2016395} that is normally 
provided in the form of a semi empirical law
\begin{eqnarray}\label{DragLaw}
\mathbf{F}_{drag}= \beta (\mathbf{u}_{f}-\mathbf{u}_{p}),\\
\beta=\beta(\mathbf{u}_{f}-\mathbf{u}_{p},\rho_f,\rho_p,d_p,A_p,\mu_f
, \epsilon),
\end{eqnarray}
with $\mathbf{u}_{f}, \, \mathbf{u}_{p}$ the fluid and particle 
velocity respectively,  $\rho_f,\,\rho_p$ the respective densities, $d_p,\,A_p$ 
the 
particle characteristic length and area, $\mu_f$ the fluid viscosity, and  
$\epsilon$ the porosity, defined as the ratio between the volume occupied by the 
fluid and the total volume of the CFD cell.
For the sake of generality, we took $\beta$ as described in~\cite{New1}.

During the parallel execution of XDEM, the simulation domain is geometrically decomposed in regularly fixed-size cells that are used to distribute the workload between the processes.  
Every process has global knowledge of the domain structure and decomposition, but only performs the calculation, and holds knowledge,
for the particles that belong to its sub-domain.
The load partitioning between processes has been shown to be often crucial for the parallel performance of 
XDEM~\cite{AlbanPartitioning}, and different partitioning strategies are available within the XDEM platform.
In this contribution, we investigate the parallel performance of XDEM when operated with geometrically uniform
partitions for the coupling with OpenFOAM. 
In figure~\ref{fig:case3_XDEM_scalability}, standalone XDEM executions compare the different partitioning algorithms, 
including the geometrically uniform approach, in order to assess their influence on the specific cases used in this work.

\subsection{Computational Fluid Dynamics (CFD)}

For the fluid solution, we refer to an unsteady incompressible flow through 
porous media with forcing terms arising from particle phase, as described in 
\cite{PozzettiIJMF,PozzettiICNAAM2016,PozzettiPowdermet}.
This system must fulfill the Navier-Stokes equations in the form,
\begin{eqnarray}\label{UFP}
\nabla \cdot \epsilon\mathbf{u}_f=\frac{\partial \epsilon}{\partial t} \, , 
\\
\frac{\partial \epsilon \mathbf{u}_f}{\partial t}+ \nabla \cdot (\epsilon 
\mathbf{u}_f\mathbf{u}_f)= -\epsilon \nabla p+
\mathbf{T_D}+\mathbf{F_b} +\mathbf{F_{fpi}},\\
\mathbf{T_D}=\nabla \cdot \left(\epsilon \mu_f\left( \nabla \mathbf{u}_f+ \nabla^T 
\mathbf{u}_f 
\right) \right),
\end{eqnarray}
with $\mathbf{u}_f$ the fluid velocity, $p$ the fluid pressure, $\mu_f$ the 
fluid viscosity, $\mathbf{F_b} $ a generic body force, $ \mathbf{F_{fpi}}$
the fluid-particle interaction force, that is the counterpart of 
$\mathbf{F}_{drag}$, which is here treated with the semi-implicit algorithm proposed in~\cite{ROBUSTCFDDEM}.

This set of equations is solved with the OpenFOAM libraries~\cite{OpenFOAM}, 
which parallelization is based on domain decomposition.
The CFD domain is split into sub-domains assigned to each process available at 
run time, over each of them a separate copy of the code is run.
The exchange of information between processes is performed at boundaries through 
a dedicated patch class as described in~\cite{OpenFOAM}.
In this contribution, the CFD domain is partitioned according to the
DEM domain, as described in section~\ref{sec:partitioning_strategy}. In figure~\ref{fig:case2_OF_scalability}, we compare standalone
OpenFOAM execution with different partitioning algorithm to assess the influence of the partitioning on our specific cases.

\subsection{Co-located partitions strategy for domain decomposition}\label{sec:partitioning_strategy}

\begin{figure*}[p]
\centering
\includegraphics[width=\textwidth]{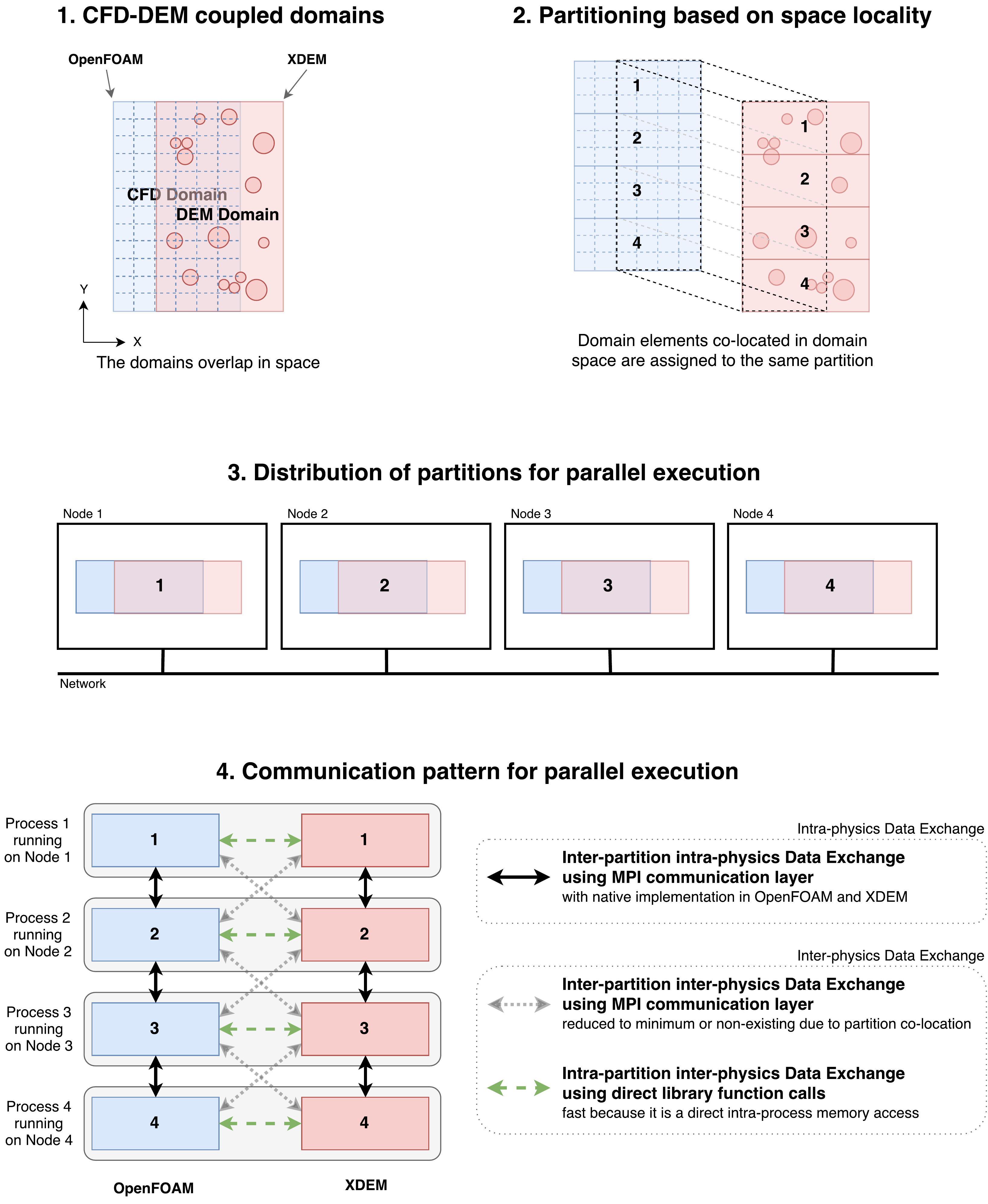}
\caption{\label{fig:partitioning_strategy} Co-located partitions strategy for a coupled CFD--DEM simulation:
1. OpenFOAM and XDEM domains are overlapping. 
2. Domain elements co-located in the simulation space are assigned to the same partitions.
3. Partitions are distributed to the computing nodes.
4. The parallel execution benefits from the resulting communication pattern.}
\end{figure*}

Since CFD--DEM couplings based on volume averaging technique only consider local exchanges of mass, momentum, and energy, 
the Lagrangian (DEM) entity will be affected only from the fluid characteristic of an area close to its center. 
At the same time, the fluid flow will be only locally interested by the action of particles present in the local region.

In this contribution, we are taking advantage of theses characteristics to
design an optimized partitioning strategy for our CFD--DEM simulation. 
This partitioning strategy is illustrated on figure~\ref{fig:partitioning_strategy}:
\begin{itemize}
  \item[1.] In our problem, the CFD (OpenFOAM) and DEM (XDEM) domains are overlapping, totally or partially. 
  As stated earlier, interactions between the fluid and the particles happen for objects closely located in the simulation space.
  \item[2.] In consequence, our partitioning strategy considers the 2 domains together and its goal is to maintain 
  the locality of the objects in the simulation. This means the domain elements co-located in the simulation space 
  are assigned to the same process. In this way, the partition $i$ of the CFD and the partition $i$ of the DEM domain will be coinciding.
  \item[3.] For the parallel execution, partitions are distributed to the computing nodes. It is important to make sure 
  that partition $i$ of the CFD domain and partition $i$ of the DEM domain are actually located in the computing node, or even better, in the same MPI process.
  \item[4.] With this partitioning strategy, the parallel execution benefits from the resulting communication pattern. 
  
  \textbf{Inter-partition intra-physics data exchanges}, i.e. the same that occur in a standalone execution 
  of OpenFOAM or XDEM, are performed with the MPI communication layer using native implementation.
  This ensures good portability of the approach as the intra-physics communication implementation is not changed.
  
  \textbf{Intra-partition inter-physics data exchanges} are implemented with simple read/write operation in memory 
  or direct library function calls because the two partitions share the same intra-process memory space. 
  These data exchanges, which represent the majority of the data exchange due to the co-located partitions strategy, 
  are much faster than the ones based on traditional communication layer like MPI.
  
  \textbf{Inter-partition inter-physics data exchanges} are achieved using the MPI communication layer. 
  Thanks to the co-located partitions strategy, they are reduced to a minimum, or non-existent if the partitions are perfectly aligned.
\end{itemize}

As a result, our approach has many advantages compared to the other work in the literature (cf section~\ref{Introduction}).
Firstly, we rely on the distribution of the simulation domain among the computing nodes. 
Every partition of the computational domain keeps only its local data, 
and therefore reduces memory consumption. 
This is important as it allows simulating large-scale problems that would not fit in the memory of a single computing node.
Secondly, by running the CFD and DEM code in the same process (by linking the two libraries into one executable), 
the local inter-physics data exchanges can be performed by direct memory accesses 
and avoid the overhead of an additional communication layer.
Thanks to our partitioning strategy that imposes the co-location constraint on the CFD and DEM sub-domains,
we ensure that the majority of the inter-physics data exchange will occur in that way.
These advantages are meant to reduce the overhead of a parallel execution and to
simplify the structure of the interface between the different codes, that will 
be identical for parallel and sequential execution.

In this study, we implement the proposed strategy by using a simple partitioning algorithm that enforces the co-location constraint.
This algorithm controls the partitioning of both the CFD and DEM domain.
For the sake of simplicity, the global simulation space is here partitioned with a uniform strategy based on a geometric decomposition.
This consists in splitting the domain into regions of similar 
volume making the spatial volume of the domains, that are assigned to each process, balanced.
In this way, every process will handle a similar volume of the computational domain holding 
data and performing the computation for the CFD and DEM part occupying the specific spatial region.
While this strategy is not optimal for a generic case, it greatly simplifies the implementation and the analysis of the results.

The result of this partitioning is then enforced into each software bypassing any internal they may use.
For OpenFOAM, this is achieved using the \emph{manual} decomposition method which allows using a predefined partitioning defined in a text file.
In this way, our partitioning algorithm considers the global simulation space of two physics software as a whole, 
not differencing if a point in space belongs to the CFD mesh or the DEM domain.
This strategy is significantly different from what proposed in previous works~\cite{SediFoam}, 
where the partitioning of CFD and DEM domains was aiming to optimize the load-balance of each standalone software,
rather than the communication cost of the coupled execution.

%But the main purpose of the current work is to test the validity of our partitioning strategy about co-locating partitions from CFD and DEM software within the same process.

\section{Experimental results}

\subsection{Experimental methodology}

In order to assess the validity of our approach and evaluate the scalability of the proposed strategy, 
we have setup and executed three benchmarks.

The first benchmark, \textit{One particle traveling across a processes boundary} presented in section~\ref{sec:case1},
tests the equivalence of the results between sequential run and parallel run with the current parallelization strategy. 
This is done to assess the validity of our approach and implementation 
by checking the continuity of the results when the particle travels across process boundary
even without inter-partition inter-physics communication.

The second benchmark, \textit{One million particles in two million cells} proposed in section~\ref{sec:case2}, 
investigates the effects of an heavy inter-physics communication on the performance of a standalone software.
This is done to show how the proposed parallelization strategy allows maintaining the performance of the original single solver 
in the presence of an heavy inter-physics communication even with hundreds of processes.

The third benchmark, \textit{Ten Million Particles in one Million cells} described in section~\ref{sec:case3},
studies the parallel performance of a coupled solution in case of an heavy coupled case. 
This is done to show how our solution can handle highly costly simulations 
and at the same time allows to resolve a main issue underlined in the literature linked to the inter-physics communication. 

\medskip

The experiments were carried out using the $Iris$ cluster of the University of Luxembourg~\cite{VBCG_HPCS14}
which provides 168 computing nodes for a total of 4704 cores.
The nodes used in this study  feature a total a 128 GB of memory 
and have two Intel Xeon E5-2680 v4 processors running at 2.4 GHz, that is to say a total of 28 cores per node.
The nodes are connected through a fast low-latency EDR InfiniBand (100Gb/s) network organized in a fat tree. 
We used OpenFOAM-Extend 3.2 and XDEM version b6e12a86, both compiled with Intel Compiler 2016.1.150
and parallel executions were performed using Intel MPI 5.1.2.150 over the InfiniBand network.

To ensure the stability of the measurement, the nodes were reserved for an exclusive access. 
Additionally, each performance value reported in this section is the average of at least hundred of measurements.
The standard deviation showed no significant variation in the results.

\subsection{One particle traveling across a process boundary}\label{sec:case1}

\begin{figure*}[htbp]
\centering
	\includegraphics[width=1\textwidth]{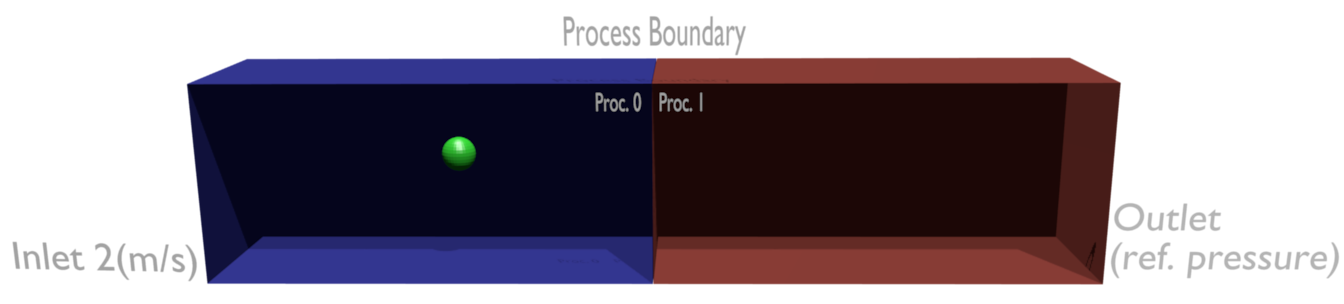}
	\caption{\label{fig:case1_setup} One particle traveling across process boundary: Setup.}
\end{figure*}

\begin{figure*}[htbp]
\centering
\subfloat[]{\label{PartCharOP}  \includegraphics[width=1\textwidth]{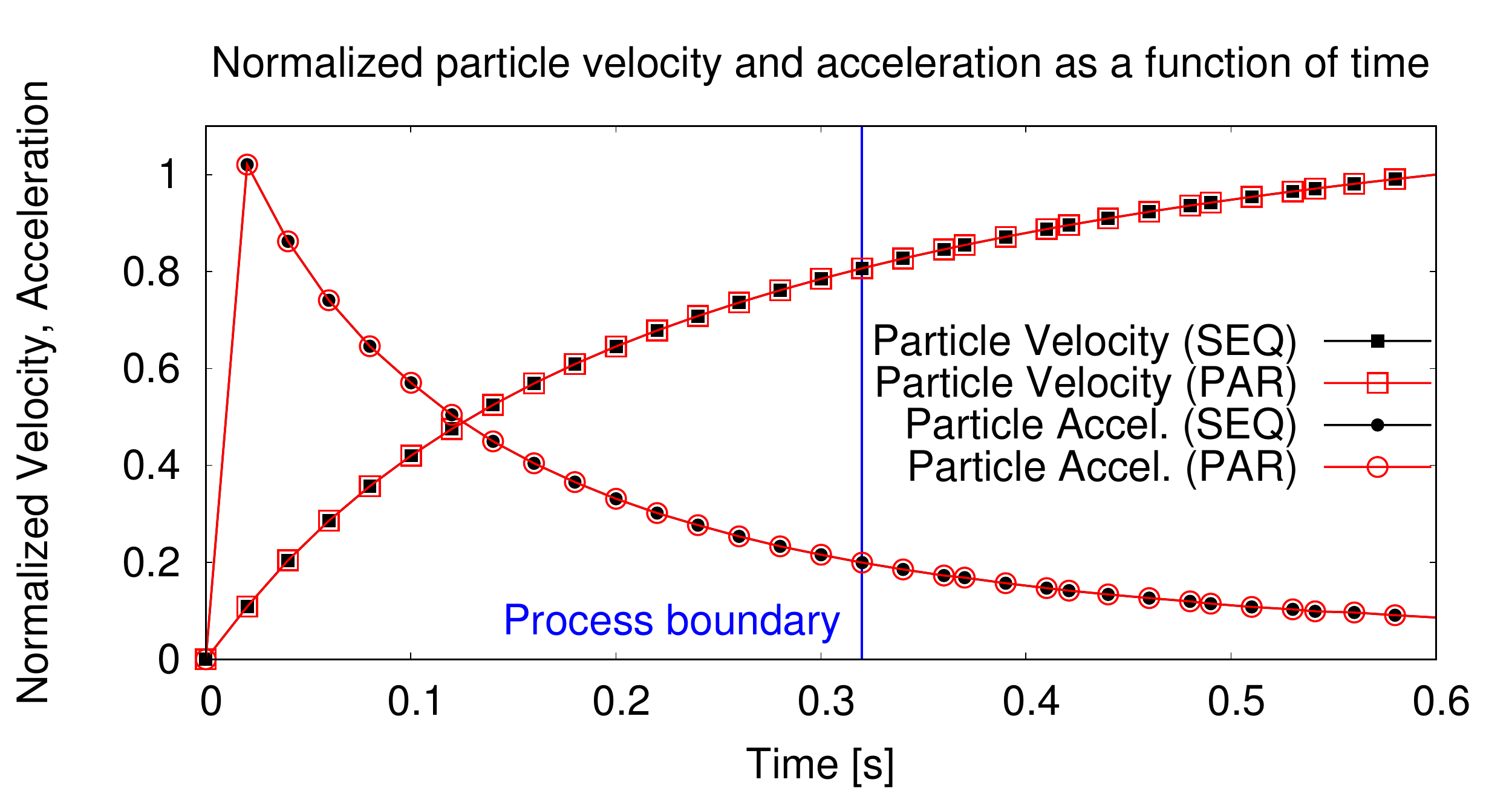} }\\
\subfloat[]{\label{FluidFieldOP}\includegraphics[width=1\textwidth]{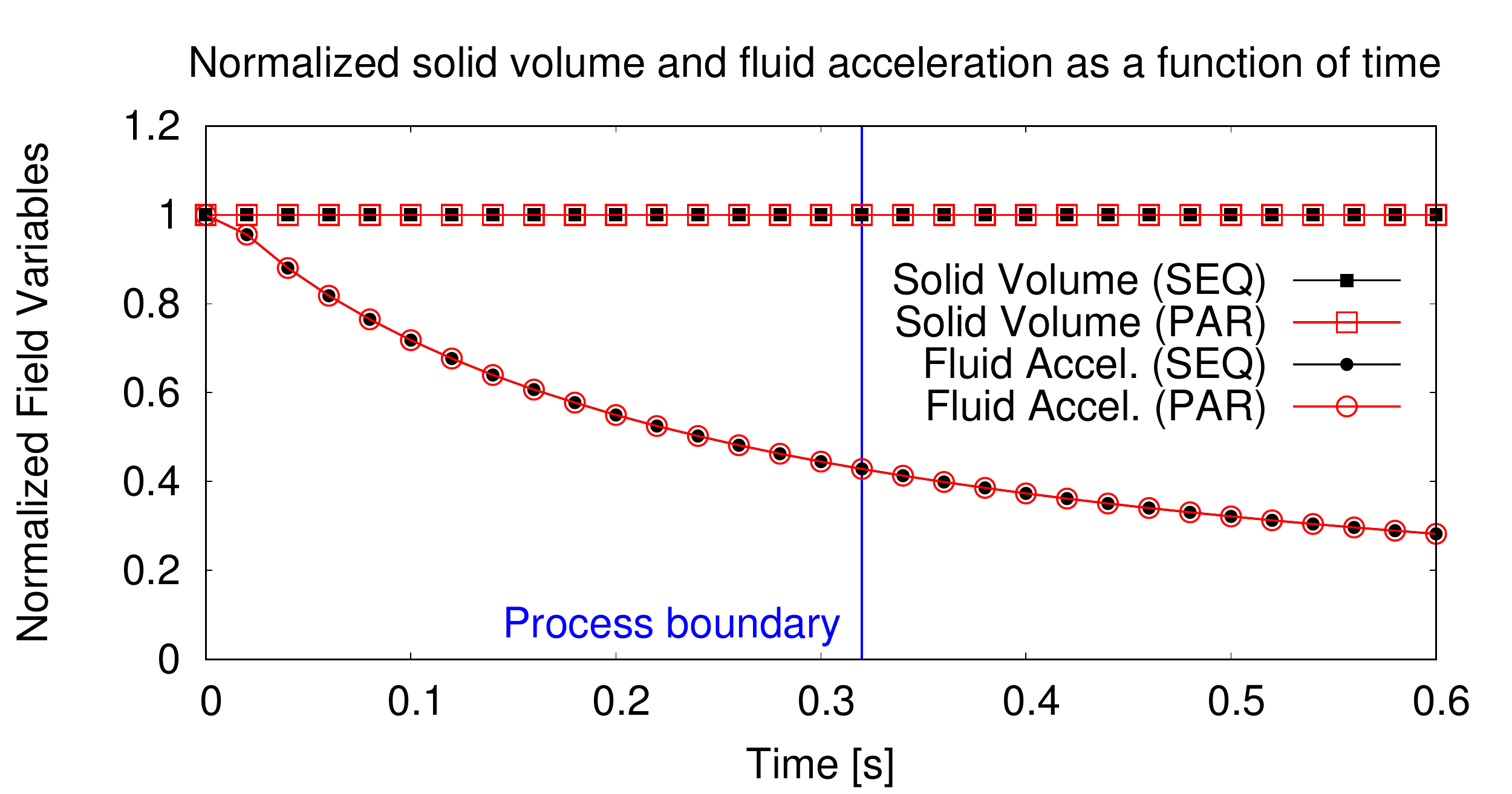}}
\caption{\label{fig:case1_plots}{
One particle traveling across process boundary:
comparison between the sequential (SEQ) and parallel (PAR) results for CFD and DEM 
variables. Continuity across the process boundary can be observed.}}
\end{figure*}

A first basic test is proposed, featuring a particle traveling 
across a boundary between processes.
This is done in order to test the equivalence of the results obtained in 
sequential and parallel execution.
The particle, initially at rest within the domain assigned to process $0$, is 
accelerated by the fluid according to the law of equation \ref{DragLaw}. 
The resulting drag force pushes the particle across the boundary with process 
$1$ causing it to be transferred from the sub-domain $0$ to the sub-domain $1$.

As shown in figure~\ref{fig:case1_setup}, the boundary conditions imposed on the fluid domain are a uniform Dirichlet at the inlet,
no-slip at the wall, and reference pressure at the outlet. 
The CFD domain is a square channel of dimensions of $1m$, $0.2m$ and $0.2m$, and is discretized with $240$ identical cubic cells. 

In figure \ref{fig:case1_plots}, the normalized particle velocity and acceleration, and the normalized 
solid volume and $L_1$ norm of the fluid acceleration are proposed as a function of times.
All the quantity are normalized by dividing them by their maximum value, so that they can be displayed on the same plot.
The particle crosses the process boundary at $0.32 s$ leaving the domain assigned to process $0$ and entering into the domain 
assigned to process $1$.

It can be observed how the particle velocity and drag force are 
continuous across the processes boundary.
This shows how the information on the fluid velocity at the particle position is correctly exchanged between
CFD and DEM code in the whole domain, including the regions between boundaries.
Similarly,  the porosity and acceleration fields projected by the particle into the Eulerian grid do not suffer
discontinuities when the particle switches between processes.
In particular, it can be noticed how the results obtained with the parallel execution 
are perfectly matching the one obtained by running the code in sequential for both particle and fluid quantities.
This shows how even without keeping global knowledge of the whole computational 
domain, continuity across processes of both Eulerian and Lagrangian quantities 
can be achieved. 

This allows greatly limiting both the memory usage and the inter-process 
communication.

\subsection{One million particles in two million cells}\label{sec:case2}

\begin{figure*}[htbp]
\centering
	\includegraphics[width=0.7\textwidth]{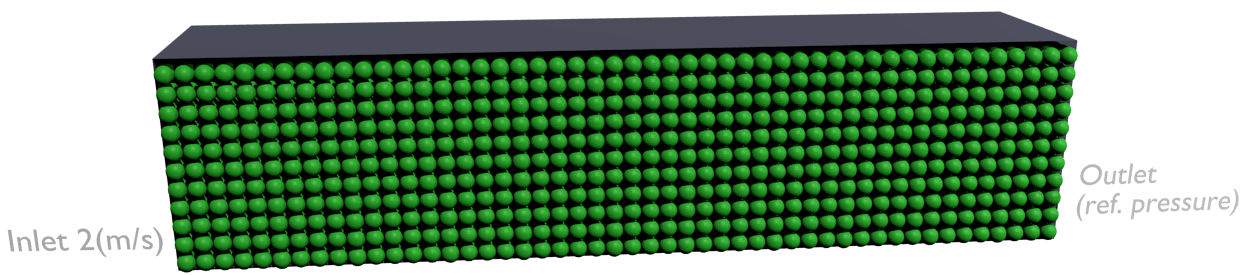}
	\caption{\label{fig:case2_setup} One million particles in two million cells: Setup.}
\end{figure*}

This second test aims to evaluate the influence of an intense intra-physics communication load on the scalability of the code.
For that, a coupling between OpenFOAM for the CFD part and a dummy DEM software, named DummyDEM, is used. 
DummyDEM, is a modified version of XDEM whose purpose is to trigger all the necessary data exchange, 
but does not perform any actually DEM computation, i.e. the particles are not moving. 

This case is similar to the one proposed in the previous section, but in order to increase the amount of data exchange
it contains a swarm of one million of particles suspended in a channel flow
discretized with two millions of identical cubical CFD cells (figure~\ref{fig:case2_setup}).
The boundary conditions imposed on the fluid domain are as in the previous test the standard inlet, wall 
and outlet conditions on respectively the channel inlet, the surrounding walls 
and the outlet section.

\subsubsection{Standalone OpenFOAM performance}

\begin{figure*}[htbp]
	\includegraphics[width=1\textwidth]{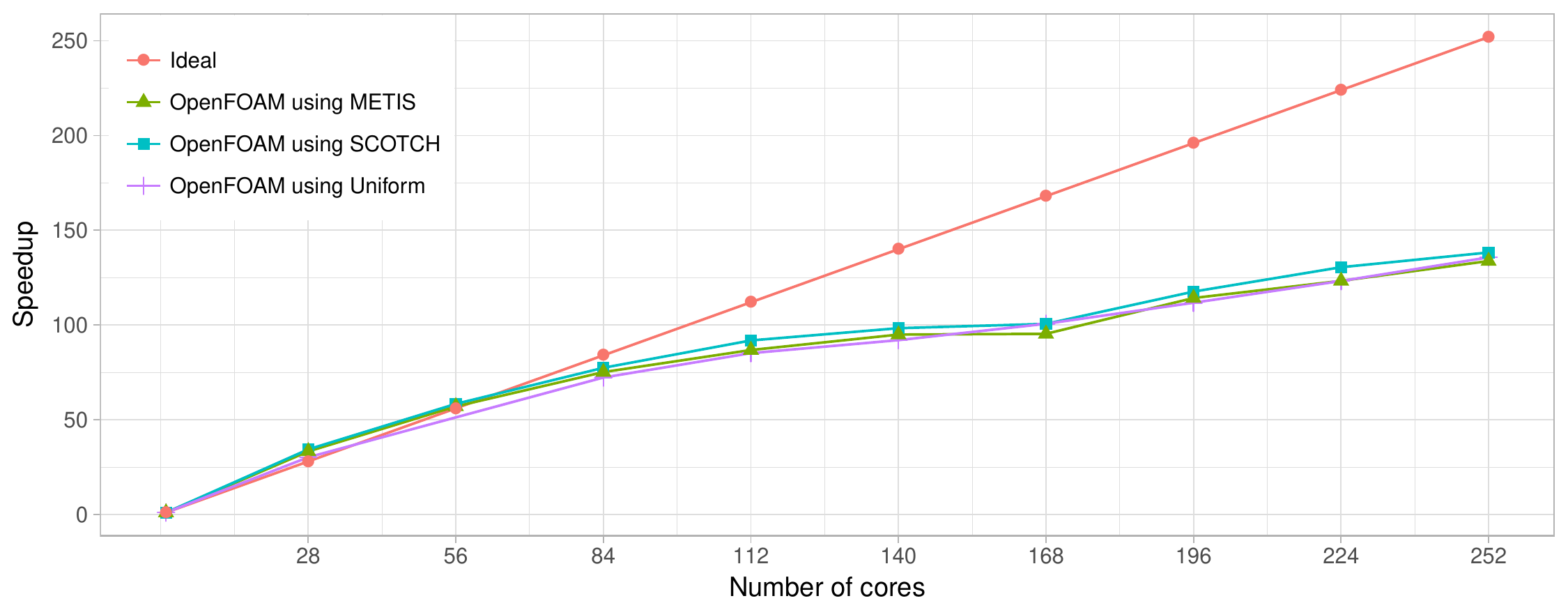}
	\caption{\label{fig:case2_OF_scalability} One million particles in two millions cells: 
		comparison between the scalability of the pure OpenFOAM partitioned 
		using SCOTCH, METIS, and uniform partitioning algorithm. Similar scalability properties
		can be observed.}
\end{figure*}

In figure~\ref{fig:case2_OF_scalability}, we first present the scalability of the standalone
OpenFOAM run obtained by decomposing the domain with the native SCOTCH and METIS partitioners,
and the uniform partitioner that is used in the coupling.
One can notice how the behavior of OpenFOAM is not highly dependent on the 
partitioner adopted. For that case, this can be explained considering that the problem is rather
uniform and therefore the optimal partitioning would be close to the uniform one.

\subsubsection{Coupled OpenFOAM--DummyDEM performance}

In a first step to understand the performance of the coupled CFD--DEM approach,
we first decided to study the behavior of the coupling in case of uniformly distributed load, 
so that the final cost is not influenced by dynamic load balancing. 
In order to obtain this, our coupling OpenFOAM--DummyDEM performs all the coupling data exchange 
but our simplified DEM implementation, DummyDEM, does not execute any DEM calculation.
As a result, in this testing coupling OpenFOAM--DummyDEM case, we have at every timestep:
\begin{itemize}
\item data exchange from CFD to DEM: the fluid velocity is used to calculate the drag force applying on the particles
\item data exchange from DEM to CFD: the distribution of the particles is used to calculate the porosity field of the CFD domain
\end{itemize}
However, DummyDEM will not integrate the particle position which will remain stationary in the simulation domain. 
Furthermore, this allows adopting the same timestep for the CFD and DEM parts, further simplifying the analysis of the results.

The resulting case is therefore equivalent to a pure CFD case, on the top of which 
a inter-physics communication is triggered. This allows studying the effects 
of the data exchange on the scalability of the software.

\begin{figure*}[htbp]
	\includegraphics[width=1\textwidth]{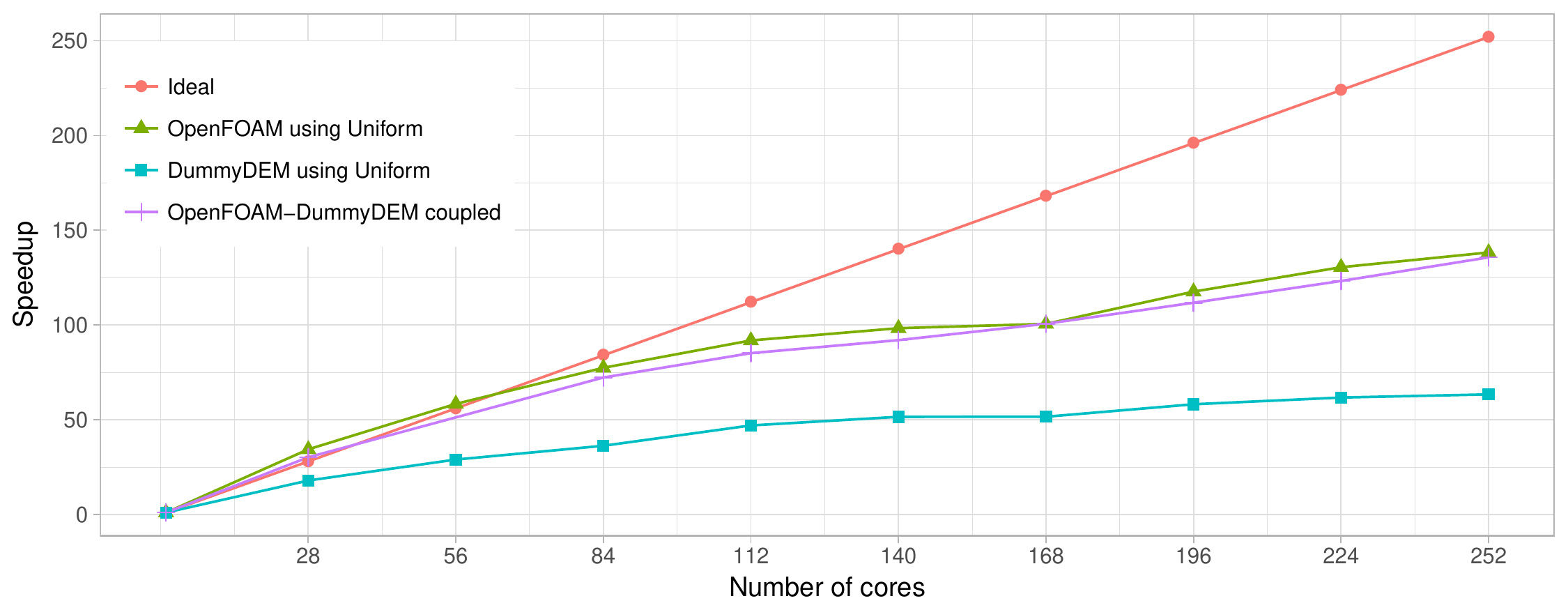}
	\caption{\label{fig:case2_XDEMOF_scalability} One million particles in two millions cells: 
        speedup of the coupled OpenFOAM--DummyDEM in comparison with standalone OpenFOAM and standalone DummyDEM.
        The scalability of the coupled OpenFOAM--DummyDEM is similar to the standalone OpenFOAM: 
        the overhead of data exchange of the coupling is negligible. 
        The poor performance of DummyDEM is due to the artificially reduced amount of computation.
    }
\end{figure*}

In figure~\ref{fig:case2_XDEMOF_scalability}, 
we evaluate the scalability of our OpenFOAM--DummyDEM coupling approach,
in comparison with standalone OpenFOAM and standalone DummyDEM execution.
The aim of this is to underline the influence of the presence of an interface 
between the Lagrangian and Eulerian domain, on the overall algorithm 
scalability.

One can observe how in this case, the parallel performance of the DummyDEM part is poor.
This is explained considering that the decision of not integrating the particle motion, 
which makes the DEM computation rather inexpensive, leading to the flat
scalability profile observed in figure~\ref{fig:case2_XDEMOF_scalability}. 
This also offers the possibility of studying  the performance of the coupled 
solver in a case where the performance of the two basic software is markedly 
different.

In this configuration, for an execution on 252 cores, we measured that 81\% of 
the execution time is spent in OpenFOAM code, 18\% in DummyDEM and less than 1\% 
doing inter-physics data exchange.
As a consequence, the performance of the OpenFOAM--DummyDEM coupling is similar to 
the one of the standalone OpenFOAM execution.
This shows how the presence of the interface between the two codes, that in this 
case is coupling one million of particles with two millions of cells,
does not affect the performance significantly. 

This can be explained considering that, in this case, the generated partitions are perfectly aligned
and, as a result, it does not introduce any inter-physics inter-process communication, 
and the most of the inter-physics data exchanges use direct memory accesses 
and avoid any communication layer.
This also proves how, by using the co-located partitions strategy, the presence of an 
intensive inter-physics data exchange does not affect the parallel 
performance of the software. This is an important step for parallel execution 
of coupled software, as it should be ensured that its performance matches 
the one of a heavy loaded standalone algorithm without an excessive overhead. 

\subsection{Ten Million Particles in one Million cells}\label{sec:case3}

\begin{figure*}[htbp]
\centering
\subfloat[]{\label{10ML}\includegraphics[width=0.8\textwidth]{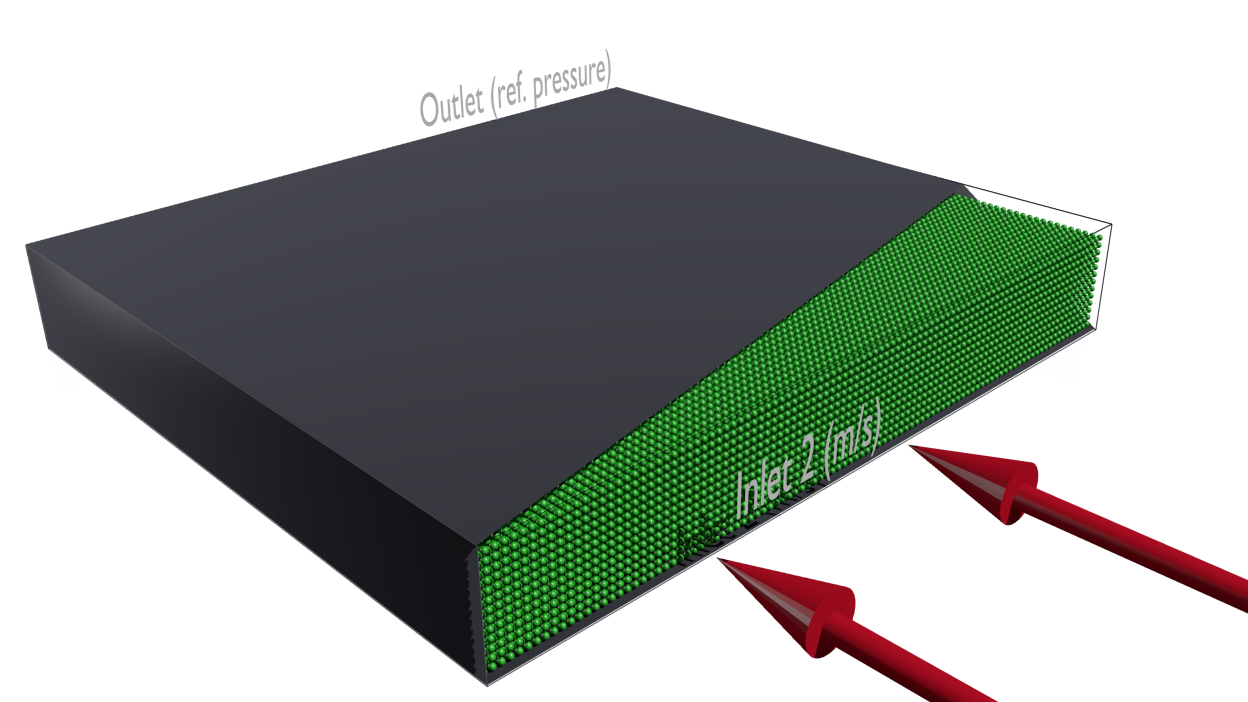}}\\
\subfloat[]{\label{10MR}\includegraphics[width=0.8\textwidth]{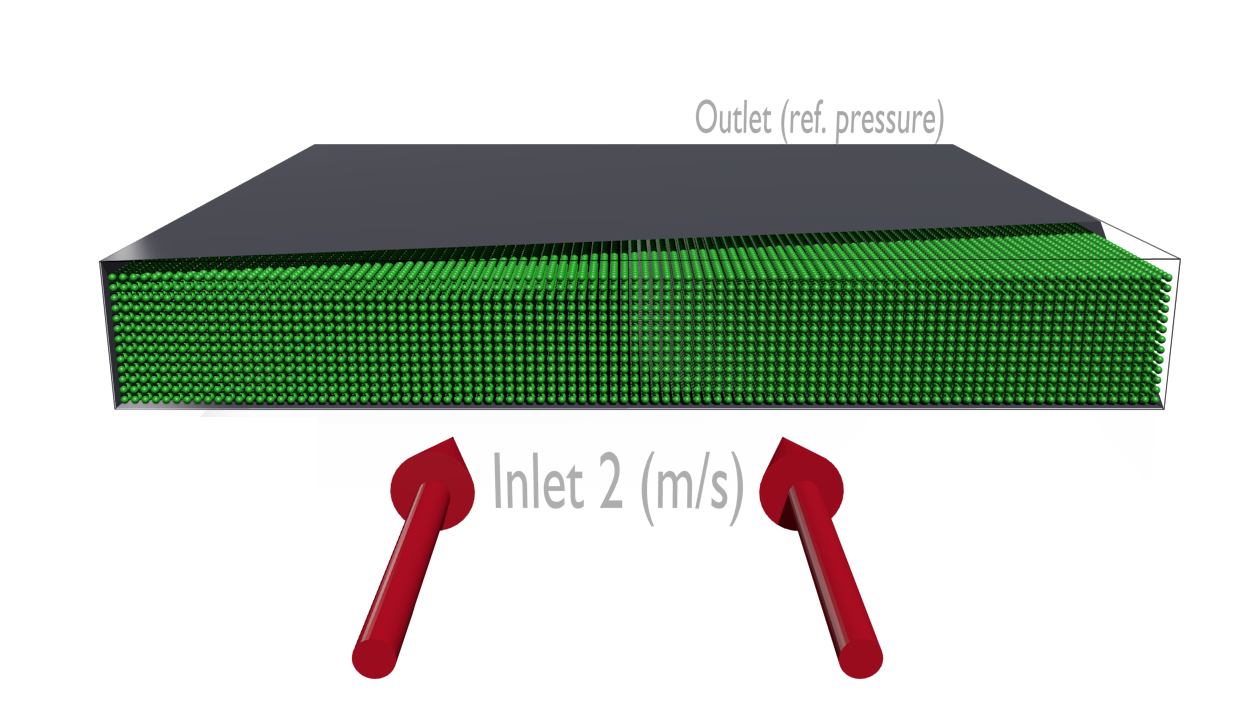}}
\caption{\label{fig:case3_setup} {Ten million particles in one million cells: Setup.}}
\end{figure*}

The third test case consists in a layered bed of ten million particles
moving in the presence of a carrier gas. This case was inspired by~\cite{SediFoam}
and chosen to show the capability of the coupling to treat a very large amount of DEM particles
while keeping the inter-physics communication cost low.

As shown in figure~\ref{fig:case3_setup}, the test case  is of the form proposed in~\cite{SediFoam}
with a domain size of 480mm x 40mm x 480 mm,
featuring a layered bed of 10 million of particles moving under the action of an incompressible flow.
The boundary conditions for the fluid solution are of constant Dirichlet at the inlet ($2 m/s$), 
non-slip at the wall boundaries and reference atmospheric pressure at the outlet.

In~\cite{SediFoam} was pointed out that with heavy coupled cases, when partitioning the two domains independently,
the inter-physics communication becomes more and more important with the rise of computing processes, reaching up
to 30\% of the whole cost of the simulation when using more than 200 computing processes.
This means that, when using an independent domain partitioning from each software, the coupled simulation will be negatively affected 
by the coupling interface, leading to a coupled execution that will perform worse than the single-code one when 
operated on a high number of computing nodes.
This behavior was also identified in~\cite{KafuiParallel} where the authors clearly showed how the coupled execution
was performing worse than both of the standalone software.

\subsubsection{Standalone XDEM and standalone OpenFOAM performance}

\begin{figure*}[htbp]
\centering
	\includegraphics[width=1\textwidth]{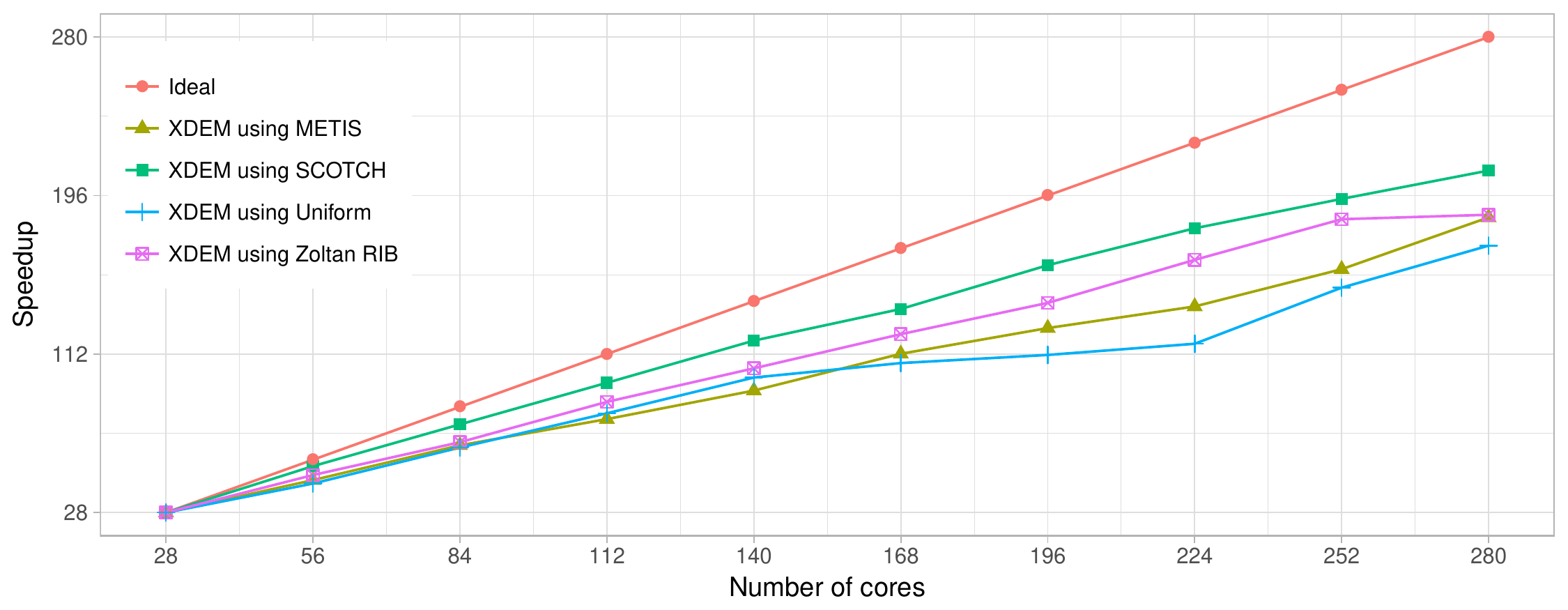}
	\caption{\label{fig:case3_XDEM_scalability}Ten million particles in one million cells: 
		Speedup of standalone XDEM for different partitioning algorithms.}
\end{figure*}

\begin{figure*}[htbp]
\centering
	\includegraphics[width=1\textwidth]{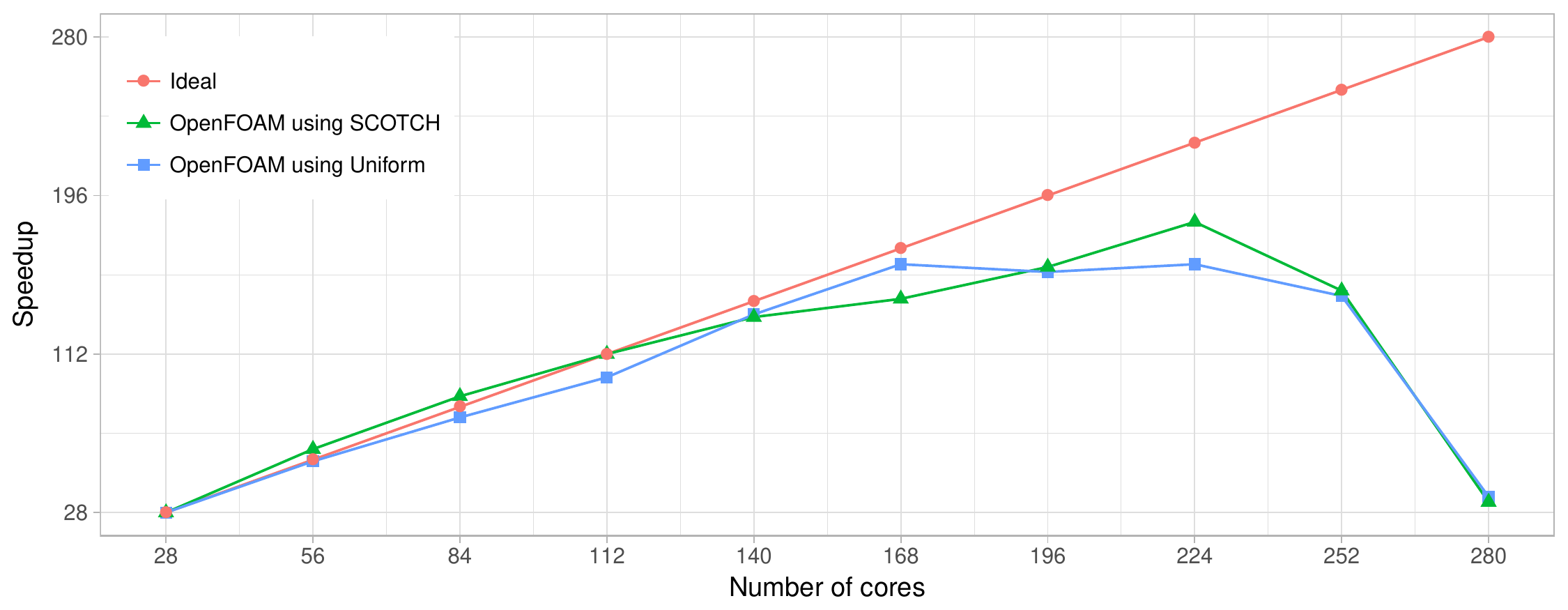}
	\caption{\label{fig:case3_OF_scalability}Ten million particles in one million cells: 
		Speedup of standalone OpenFOAM for different partitioning algorithms.}
\end{figure*}

In figure~\ref{fig:case3_XDEM_scalability}, the scalability performance of the pure DEM execution obtained with different
partitioning algorithms is proposed. One can observe how  the results are affected by the choice
partitioning algorithm, but the general behavior is rather similar. This can  be explained considering that the 
setup for this case is rather homogeneous and uniformly distributed. 

Similarly, in figure~\ref{fig:case3_OF_scalability}, the performance of the pure CFD execution obtained with a uniform partitioning 
and a classic OpenFOAM partitioner (SCOTCH) are compared. Once more, the trends  are rather similar and only minor differences can be noticed.

\subsubsection{Coupled OpenFOAM--XDEM performance}

\begin{figure*}[htbp]
\centering
	\includegraphics[width=1\textwidth]{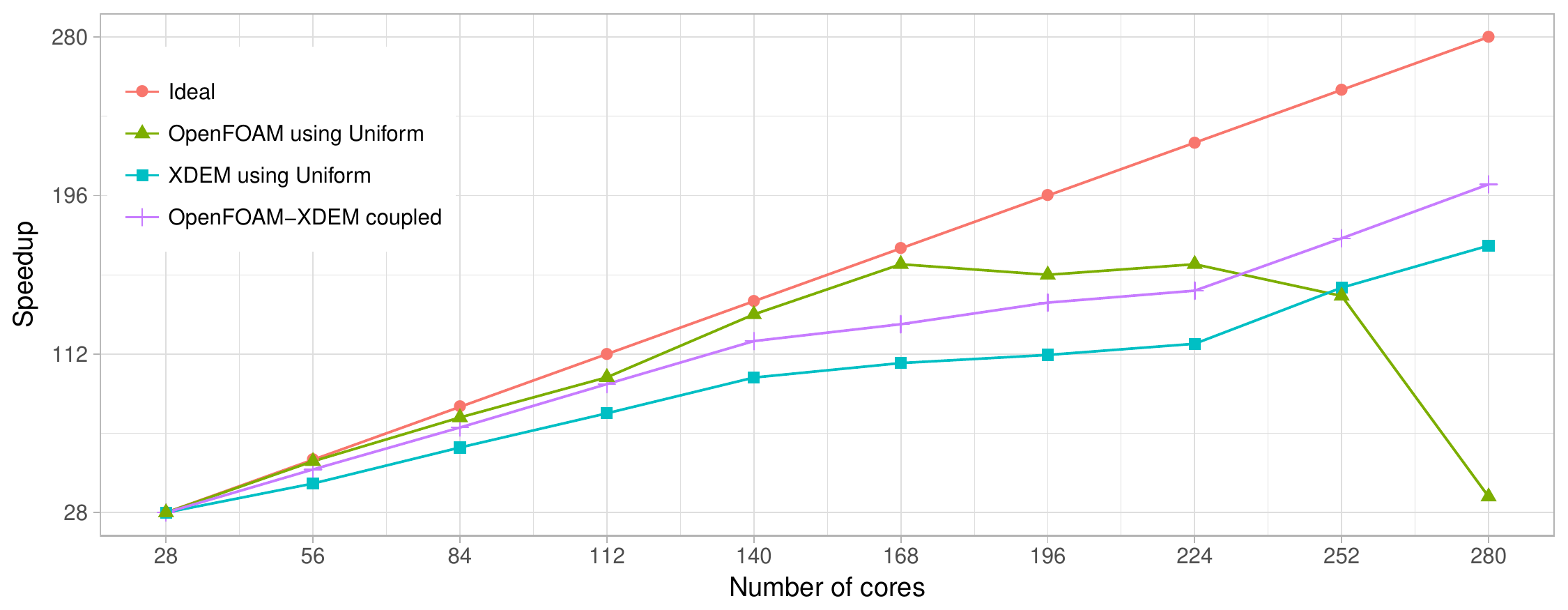}
	\caption{\label{fig:case3_XDEMOF_scalability}Ten million particles in one million cells: 
		Speedup of the coupled OpenFOAM--XDEM approach in comparison with standalone XDEM and OpenFOAM:
		The coupled execution performs better than the standalone XDEM even for more than 200 processes.}
\end{figure*}

\begin{figure*}[htbp]
\centering
	\includegraphics[width=1\textwidth]{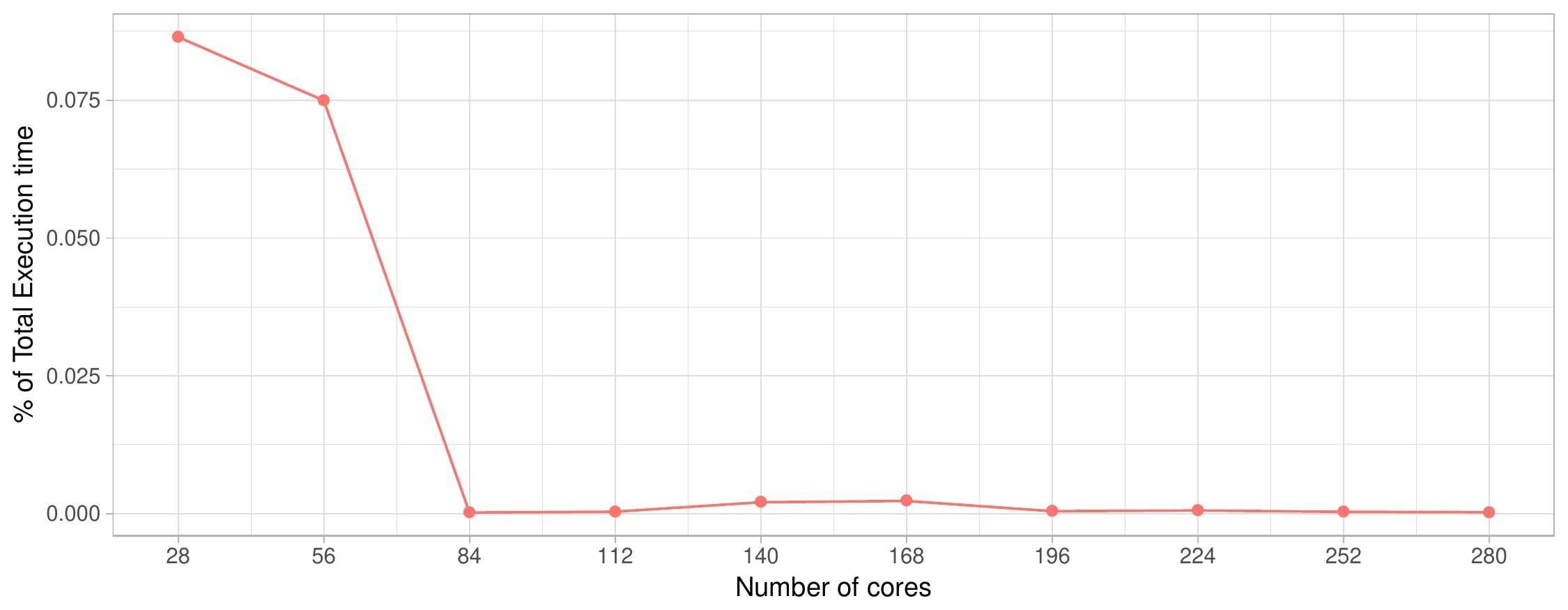}
	\caption{\label{fig:case3_XDEMOF_interphysics}Ten million particles in one million cells: 
		Percentage of time spent for inter-physics data exchange for an increasing number of processes. 
        Less than 0.1\% of the time is spent in the coupling part and the proposed solution scales well.}
\end{figure*}

In figure~\ref{fig:case3_XDEMOF_scalability}, we evaluate the performance of our coupled OpenFOAM--XDEM approach 
in comparison with standalone XDEM and standalone CFD, all executed with a uniform partitioning. 
It is interesting to notice how the parallel execution of the coupled code is not worse than the one of the single code when operating 
with more than 200 processes. 
This can be explained by the repartition of the computational cost: more than 92\% is spent in the XDEM part, less than 8\% in OpenFOAM 
and the inter-physics data exchanges represented less than 0.1\% as shown on figure~\ref{fig:case3_XDEMOF_interphysics}. 
It appears that, in this case, the cost of the coupling is negligible and the behavior of the coupled solution is driven 
by the most computation intensive part of the code, namely XDEM.

This represents a major improvement compared to previous work. 
In particular, instead of rising as it was happening for~\cite{SediFoam}, the inter-physics data exchange reduces in magnitude when using our co-located partitions
strategy. This can be explained considering that  
no inter-process communication is involved, and the amount of the data exchanged per each process reduces. In this way, instead of becoming one of the limiting
factors for the scalability of a coupled execution, this communication benefits from being executed in parallel when a co-located partitions strategy is 
adopted.
This represents an important advantage for large-scale applications that need to run over hundreds of processes in order 
to produce results within a reasonable computational time.

\section{Discussion}

The main novelty of the current approach presented in section~\ref{sec:partitioning_strategy} is to introduce a new partitioning strategy for CFD and DEM domains.
This strategy aims to co-locate partitions from the CFD and DEM domain that will need to perform information exchange, in the same MPI process. 
This is radically different from what previously proposed in the literature, where most of the works focused on optimizing 
the load-balancing of each software independently from each other while
we focus first on reducing the inter-physics communication.

We implemented this strategy by defining a partitioning algorithm that forces CFD and DEM to be co-located.
In the cases proposed in this paper, the co-location constraint is completely fulfilled, meaning that all the information required for the inter-physics exchanges
are present within the same MPI process.
Therefore, the inter-physics inter-process communications are completely absent.
As shown in section~\ref{sec:case3}, this strategy allows overcoming the inter-physics communication bottleneck that was identified in ~\cite{SediFoam}.

Nevertheless, adopting a co-location constraint introduces a significant limitation in the partitioning flexibility of the 
two domains. 
First, an hybrid partitioning algorithm must be adopted that can optimize
the executions of the two codes.
In uniform cases, as the one proposed in this work, a uniform partitioning algorithm allows achieving reasonably good parallel performances.
On the other hand, in cases in which the DEM and CFD parts have significantly different parallel requirements (as, for instance, 
when the computational load of the DEM part is clustered in a specific region of the domain, while the CFD load is uniform)
a better partitioning algorithm that optimizes the coupled execution must be designed.

We want to point out how, for more complex partitioning algorithms, a perfect fulfillment of the co-location constraint might be difficult
or not possible. In this cases, one might consider relaxing the co-location constraint allowing the partitions to be non-perfectly coinciding.
This will require the introduction of an inter-physics inter-partition communication layer that will cause a communication overhead with respect to the 
cases here presented. This cost will be null for perfectly coinciding partitions, and maximum for completely non-coinciding partitions. 
Therefore, an optimal partitioning algorithm for CFD--DEM couplings must consider both the cost associated with intra and inter-physics communication 
in order not to be excessively penalized by any bottleneck.

% Some classical partitioning algorithms like METIS or SCOTCH can be probably be used to achieved this difficult task
% if the graph describing the load of the application is constructed properly. 
% One challenge is to find the proper weight of the graph nodes that will correctly reflect the computational load of each software.

Alternatively, in several cases, it might not be always necessary to have such a complex hybrid partitioning algorithm.
When one of the software is clearly dominating in term of computational load,
it could be sufficient to partition its domain with one of its native algorithms, 
and decide the partitions of the second software to match the decision of the dominant one, 
respecting the co-location principle proposed in this work.

% The second limitation is the alignment of the partitions between the DEM and CFD domains.
% Our testcases were designed to avoid this issue and ensure that the frontiers of the partitions between DEM and CFD always align properly.
% For more complex partitioning strategies, a perfect fulfillment of the alignment constraint might be difficult or not possible. 
% In this cases, one might consider to relax the alignment constraint allowing the partitions to be non-perfectly aligned.
% This will require the introduction of an inter-physics inter-partition communication layer that will cause an additional overhead. 
% 
% Once again, we could design an advanced partitioning algorithm that would consider those constraints 
% and favor cuts where the partition can be aligned.
% On the other hand, the case itself could be also designed, by adapting the CFD mesh of the DEM grid, to help the partitioner in this task.

%In the end, the partition algorithm should consider the global simulation space and take into account both the costs associated to intra and inter-physics communication.

\section{Conclusion}

A partitioning strategy for CFD--DEM couplings has been investigated.
It consists of the usage of co-located partitions for the CFD and the 
DEM domains. This 
allows performing the bulk information exchange between the two codes locally,
reducing the global communication time and in particular minimizing the 
inter-physics parallel communication.
This strategy was implemented by using a simple uniform partitioning algorithm that forces
the partitions to be co-located.

Three benchmark cases have been introduced in order to assess the consistency 
and performances of the proposed strategy.
Experimental results were carried out using the coupling between the XDEM framework and the OpenFOAM libraries.
The validity of the results have been assessed by comparing sequential and parallel executions,
and scalability performance over hundreds of processes has been reported 
showing that less than 0.1\% of the time is used for inter-physics data exchange.

The main advantage of the proposed strategy consists in the reduction of the cost associated 
with the inter-physics communication, that is a fundamental step toward
the large-scale computation, as it allows maintaining good parallel performances 
when operating with hundreds of computing processes.
This results proved how the inter-physics communication must be taken into account 
when choosing a partitioning algorithm for CFD--DEM couplings in order to avoid an important communication bottleneck.

In order to improve the applicability our work to more generic scenarios, 
an enhanced flexibility in the domain partitioning can be provided.
For this, dedicated partitioning algorithms must be designed 
to find the good trade-off between a good load-balancing within each software, their own communication requirements 
and the volume of inter-physics data exchange. This will be studied in future works.

% This is meant to provide a reference and benchmarks for future works in this 
% field featuring improved or different 
% parallelization strategies

\section*{Acknowledgement}
This research is in the framework of the project DigitalTwin, supported by the 
programme “Investissement pour la comp\'etitivit\'e et emploi” - European Regional Development Fund (Grant agreement: 2016-01-002-06).

The experiments presented in this paper were carried out
using the HPC facilities of the University of 
Luxembourg\footnote{\texttt{http://hpc.uni.lu}}.
%~\cite{VBCG_HPCS14}.

\bibliographystyle{plain}%

\end{document}